\documentclass[11pt,letterpaper]{article}
\usepackage{systeme} 
\usepackage[utf8]{inputenc}
\usepackage{multirow} 
\usepackage{booktabs} 
\usepackage{array} 
\usepackage[english]{babel}
\usepackage{enumitem} 
\usepackage{amsmath}
\usepackage{amsfonts}
\usepackage{amssymb}
\usepackage{graphicx}
\usepackage[small,bf]{caption} 
\usepackage[left=3cm,right=3cm,top=2cm,bottom=3cm]{geometry}
\usepackage{framed} 
\usepackage{color} 
\usepackage{wrapfig}\definecolor{shadecolor}{RGB}{220,220,220} 

\author{Jorge Pinochet}
\title{\textbf{Hawking for everyone: Commemorating half a century of an unfinished scientific revolution}}
\begin{document}

\author{Jorge Pinochet$^{*}$\\ \\
 \small{$^{*}$\textit{Facultad de Ciencias Básicas, Departamento de Física. }}\\
  \small{\textit{Centro de Investigación en Educación (CIE-UMCE),}}\\
 \small{\textit{Núcleo Pensamiento Computacional y Educación para el Desarrollo Sostenible (NuCES).}}\\
 \small{\textit{Universidad Metropolitana de Ciencias de la Educación,}}\\
 \small{\textit{Av. José Pedro Alessandri 774, Ñuñoa, Santiago, Chile.}}\\
 \small{e-mail: jorge.pinochet@umce.cl}\\}

\date{}
\maketitle

\begin{center}\rule{0.9\textwidth}{0.1mm} \end{center}
\begin{abstract}
\noindent This year marks half a century since Stephen Hawking made his greatest scientific discovery by theoretically proving that “black holes ain’t so black”, as they behave like hot bodies with an absolute temperature that depends inversely on their mass. This discovery is expressed by a simple and elegant equation known as the Hawking temperature. The best way to commemorate this great scientific event is by bringing it to a wide audience. The simplest and most transparent and intuitive tool to achieve this goal is dimensional analysis. The objective of this work is to use this tool to derive the Hawking equation, reveal its meaning, and explore its main physical consequences.\\


\begin{center}\rule{0.9\textwidth}{0.1mm} \end{center}
\end{abstract}

\maketitle

\section{Introduction}

This year marks half a century since Stephen Hawking revolutionised physics by proposing, in 1974, that “black holes ain’t so black” (see Fig. 1). Specifically, Hawking theoretically demonstrated that a black hole emits thermal radiation as if it were a hot body with an absolute temperature $T_{H}$, known as the \textit{Hawking temperature} [1,2], which can be calculated by the equation:

\begin{equation} 
T_{H} = \frac{\hbar c^{3}}{8\pi kG M_{BH}},
\end{equation}

where $M_{BH}$ is the mass of the black hole, and $\hbar,c,k,G$ are fundamental physical constants. Despite the scientific relevance of Hawking temperature, and the time that has passed since it was formulated, it continues to be unknown, and at best misunderstood, by the vast majority of people.\\

The best way to commemorate the 50th anniversary of Hawking's discovery is to bring it to a wide audience. The simplest and most transparent and intuitive tool for this task is dimensional analysis [3], the application of which requires a minimal knowledge of physics and mathematics. In this work, we will use dimensional analysis to carry out an original derivation of the Hawking temperature. Following this, we will explore the meaning of this equation and its main physical consequences, namely that black holes evaporate and have entropy.\\

This manuscript may be useful to gifted high school students, physics teachers, scientists and engineers, and, more generally, to any educated person who is interested in learning about Stephen Hawking's most important scientific contribution.

\begin{figure}[h]
  \centering
    \includegraphics[width=0.3\textwidth]{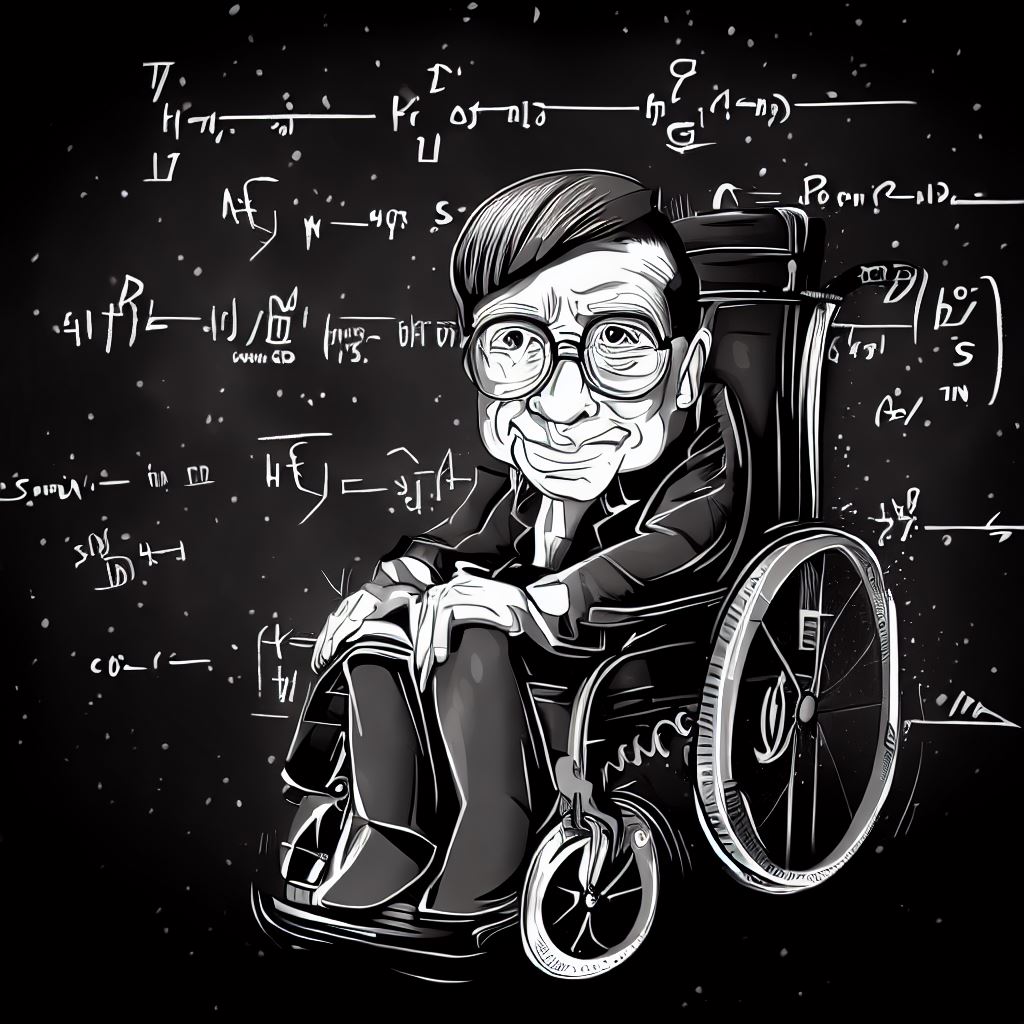}
  \caption{Stephen Hawking, born in 1942, was a renowned British theoretical physicist known for his contributions to cosmology and black hole physics. His greatest scientific discovery was the equation that established that black holes behave as hot bodies with an absolute temperature that depends inversely on their mass. Despite facing progressive motor paralysis caused by Lou Gehrig's disease, Hawking continued to research and communicate his revolutionary ideas until his death in 2018.}
\end{figure}

\section{Black holes and Hawking's discovery}

Black holes are the most extreme prediction of general relativity, which is the theory of gravity proposed by Einstein to refine and extend Newton's law of gravitation [4]. A black hole is a region of space where there is such a high concentration of matter that nothing can escape its powerful gravity, not even light [5]. The simplest of these objects is the so-called static black hole, whose mathematical description depends only on its mass. Eq. (1) applies to these objects, so we will focus on them here.\\

Fig. 2 shows an intuitive representation of a static black hole [5]. The mass of this object is concentrated in a point region called a \textit{singularity}, located in the centre of the \textit{event horizon}, or simply \textit{horizon}, which is a spherical surface that defines the outer limit of the black hole. Although the horizon has no material existence, it can be imagined as a unidirectional membrane that only allows matter or energy to flow inward [7]. The radius of the horizon, called the \textit{gravitational radius} (also know as \textit{Schwarzschild radius}), is calculated as [6]:

\begin{equation} 
R_{g} = \frac{2GM_{BH}}{c^{2}},
\end{equation}

where $G$ is the gravitational constant, $c$ is the speed of light in a vacuum, and $M_{BH}$ is the mass of the black hole, that is, the mass confined in the singularity. Since nothing can cross the horizon to the outside, this region appears completely black. If the description provided by general relativity is correct, we see that the laws of thermodynamics ensure that the temperature of the horizon is strictly zero; otherwise, the black hole would emit thermal radiation and would not be black.\\

\begin{figure}[h]
  \centering
    \includegraphics[width=0.2\textwidth]{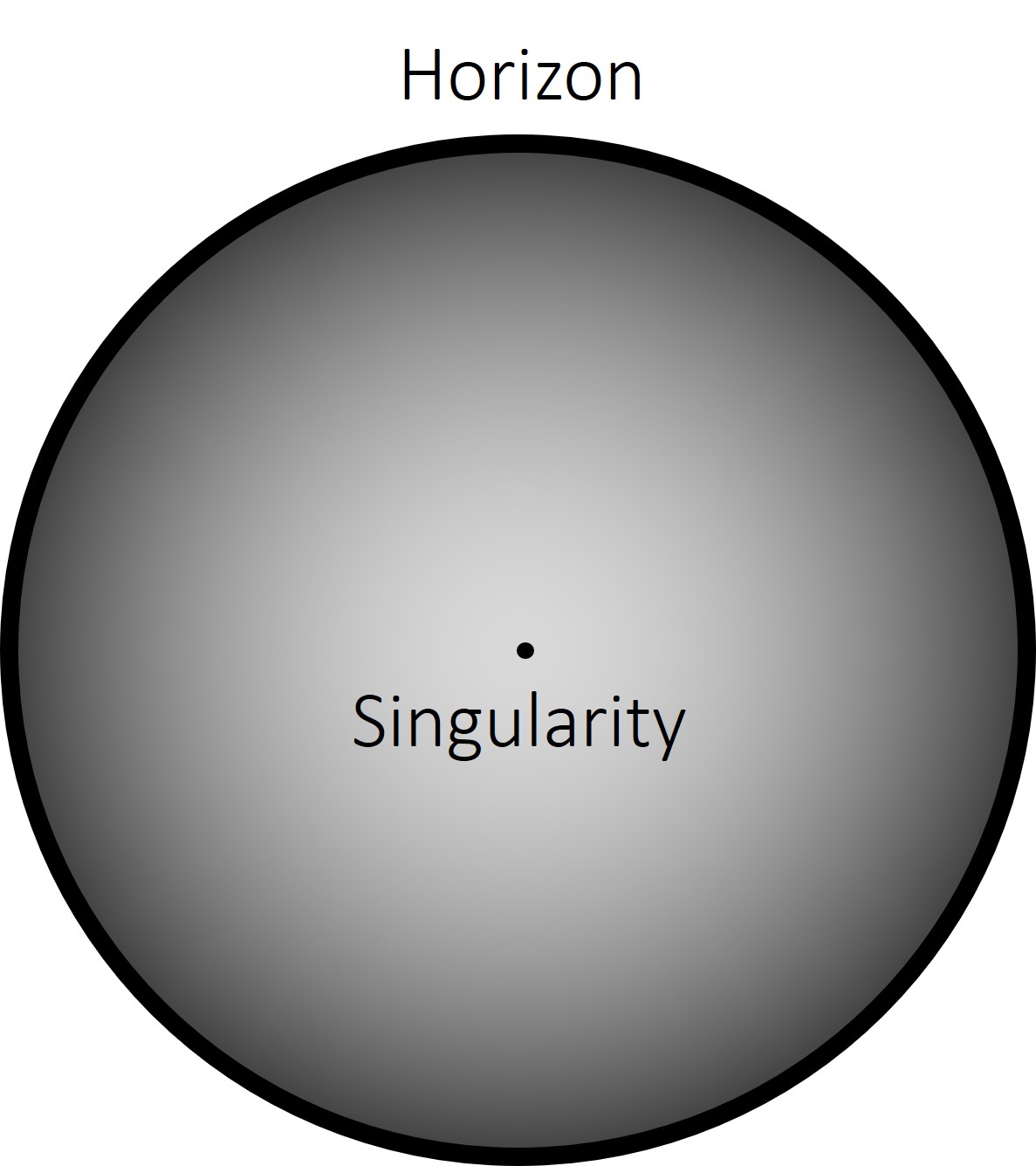}
  \caption{Structure of a static black hole.}
\end{figure}

Broadly speaking, this is the classic black hole paradigm that emerges from general relativity. In 1974, when Hawking entered the scene, no one questioned this paradigm. However, Hawking dared to question it, proving that “black holes ain’t so black”. Specifically, Hawking proved that the horizon behaves as a hot body with an absolute temperature $T_{H}$, the Hawking temperature [1,2,8], which is proportional to the gravity at the horizon.

\begin{figure}[h]
  \centering
    \includegraphics[width=0.6\textwidth]{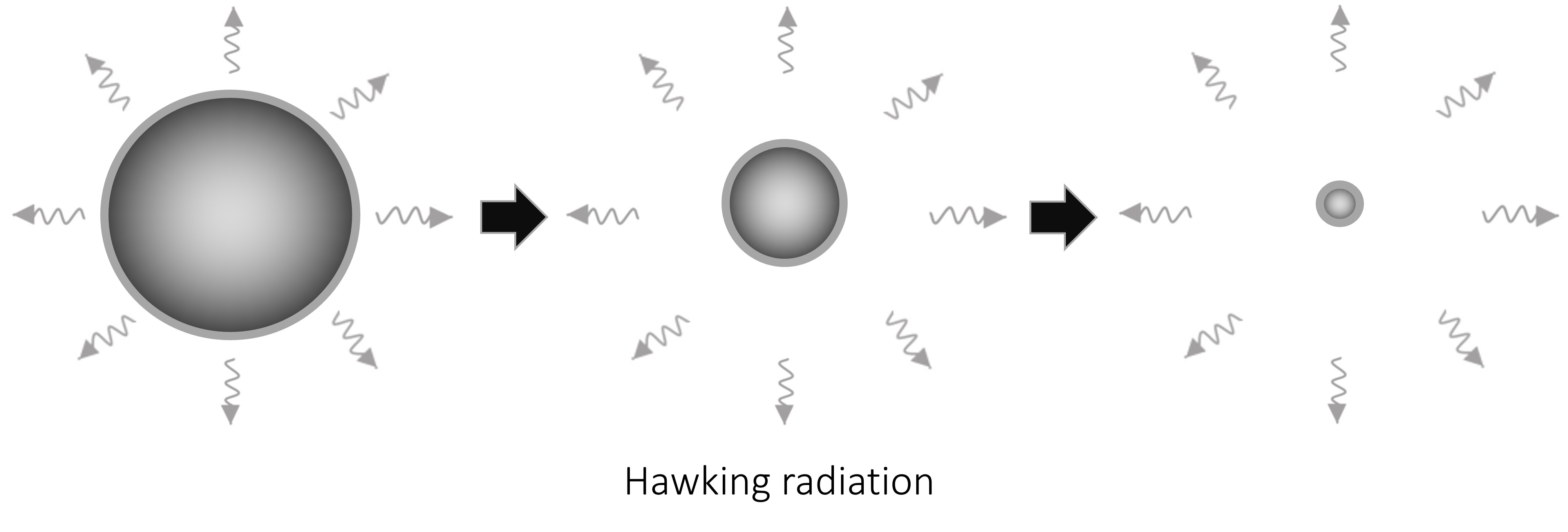}
  \caption{Black holes emit thermal radiation, have entropy, and gradually evaporate.}
\end{figure}

\begin{table}[htbp] 
\begin{center}
\caption{Theories and physical constants involved in the Hawking temperature}
\resizebox{0.8\textwidth}{!} {
\begin{tabular}{l l l} 
\toprule
\textbf{Theory} & \textbf{Characteristic constant} & \textbf{Name} \\
\midrule
\multirow{2}{3cm}{General relativity} & $G = 6.67 \times 10^{-11} N\cdot m^{2} \cdot kg^{-2}$ & Gravitational constant \\ 

& $c=3\times 10^{8} m\cdot s^{-1}$ & Speed of light in vacuum \\ 

Quantum mechanics & $\hbar = 1.05\times 10^{-34} kg\cdot m^{2} \cdot s^{-1}$ & Planck constant  \\ 

Thermodynamics & $k = 1.38\times 10^{-23} kg \cdot m^{2} \cdot K^{-1}$ & Boltzmann constant  \\
\midrule
\end{tabular}
}
\label{Theories and physical constants involved in the Hawking temperature}
\end{center}
\end{table}

As a consequence, the horizon emits thermal radiation in all directions, the so-called \textit{Hawking radiation} (Fig. 3). According to Einstein's mass-energy equivalence, the energy radiated by the black hole leads to a gradual reduction of $M_{BH}$ and $R_{S}$ through a process called \textit{evaporation} [7,9]. Furthermore, from the laws of thermodynamics, this implies that a black hole has entropy, which turns out to be proportional to the area of the horizon. From Eq. (2), this area is calculated as:

\begin{equation} 
A_{BH} = 4\pi R_{g}^{2} = \frac{16\pi G^{2}M_{BH}^{2}}{c^{4}}.
\end{equation}

To reach these conclusions, Hawking (partially) combined general relativity, which describes the universe at a macroscopic level, with quantum mechanics, which describes the universe at a microscopic level, and thermodynamics, which describes thermal phenomena [2]. Table 1 shows the fundamental physical constants that characterise these theories, which will be very useful when dimensionally deriving the Hawking temperature.

\begin{table}[htbp] 
\begin{center}
\caption{SI units and their corresponding dimensions}
\resizebox{0.5\textwidth}{!} {
\begin{tabular}{l l l l} 
\toprule
\textbf{Dimension } & \textbf{Symbol} & \textbf{SI unit} & \textbf{SI symbol} \\
\midrule
Mass & $M$ & kilogram & $kg$ \\ 

Lenght & $L$ & metre & $m$ \\ 

Time & $T$ & second & $s$  \\ 

Temperature & $\Theta$ & kelvin  & $K$  \\
\midrule
\end{tabular}
}
\label{SI units and their corresponding dimensions}
\end{center}
\end{table}

To facilitate the derivation, which is presented in the next section, the constants in Table 1 are given using the basic units of the international system (SI), which allows for a direct “translation” from the units to the corresponding dimensions. Table 2 shows this translation, showing the dimensions that are relevant for our purposes along with their symbols and the associated SI units. To guide readers who are not familiar with dimensional analysis, an appendix has been included that summarises the main concepts of this tool.

\section{Derivation of the Hawking temperature}

All the physics we need for a dimensional derivation of the Hawking temperature is summarised in Tables 1 and 2, and in the idea that the description of a static black hole depends only on its mass. We will start by assuming that we do not know the Hawking equation. Obviously, our ignorance on the subject cannot be complete: in fact, to carry out a dimensional derivation, it is always necessary to start with a hypothesis about the dimensional parameters involved (variables and physical constants) [10,11].\\

To formulate our hypothesis, we must keep in mind that Hawking's great achievement was to combine general relativity with quantum mechanics and thermodynamics to demonstrate that a static black hole has an absolute temperature $T_{H}$. As shown in Table 1, this suggests that $T_{H}$ depends on the characteristic constants of these three theories, $G,c,\hbar,k$. On the other hand, we also know that a static black hole is completely characterised by its mass, $M_{BH}$. This shows us that a total of six dimensional parameters are involved, $T_H,G,c,\hbar,k,M_{BH}$, and four dimensions, $M,L,T,\Theta$. However, we can reduce these parameters to five, and thus simplify the dimensional derivation of $T_{H}$, by introducing a new variable which in astrophysics is known as the \textit{standard gravitational parameter}, and for a celestial body of mass $m$ is defined as $\mu \equiv Gm$. The usefulness of this parameter lies in the fact that in problems where gravity applies, as is the case with Hawking's discovery, the product $Gm$ appears frequently\footnote{Readers who are familiar with the fundamental theorem of dimensional analysis, the Pi theorem, will note that the use of $\mu$ is essential to this derivation. Indeed, with $p=6$ parameters ($M_{BH},T_{H},G,c,\hbar,k$) and $d=4$ dimensions ($M, L, T, \Theta$), the number of dimensionless groups or Pi groups that can be formed is $d-p=2$, which implies that the equation for $T_{H}$ remains undetermined. However, with the introduction of $\mu$, we have $d=5$ parameters, and since the number of dimensions remains $d=4$, the number of Pi groups is reduced to $d-p=1$, and the equation for $T_{H}$ is uniquely determined.}. 

\begin{table}[htbp] 
\begin{center}
\caption{Dimensional parameters, symbols and dimensions}
\resizebox{0.7\textwidth}{!} {
\begin{tabular}{l l l} 
\toprule
\textbf{Parameter} & \textbf{Symbol} & \textbf{Dimensions} \\
\midrule
Hawking temperature & $T_{H}$ & $M^{0}L^{0}T^{0}\Theta^{1}$ \\ 

Speed of light in vacuum & $c$ & $M^{0}L^{1}T^{-1}\Theta^{0}$ \\ 

Planck constant & $\hbar$ & $M^{1}L^{2}T^{-1}\Theta^{0}$ \\ 

Boltzmann constant & $k$ & $M^{1}L^{2}T^{-2}\Theta^{-1}$ \\

Standard gravitational parameter & $\mu (=GM_{BH})$ & $M^{0}L^{3}T^{-2}\Theta^{0}$  \\
\midrule
\end{tabular}
}
\label{Dimensional parameters, symbols and dimensions}
\end{center}
\end{table}

Under these conditions, if we define the standard gravitational parameter as $\mu \equiv GM_{BH}$, we can formulate our hypothesis in the form:\\ 

\textbf{\textit{Hypothesis}}: $T_{H} = f(c,\hbar, k, \mu)$.\\

The specific information needed to derive the Hawking equation is summarised in Table 3, which contains the dimensional parameters, their symbols, and the corresponding dimensions. We can rewrite our hypothesis in the form:

\begin{equation} 
T_{H} = \Pi c^{x_{1}} \hbar^{x_{2}}k^{x_{3}}\mu^{x_{4}},
\end{equation}

where $\Pi$ is an indeterminate dimensionless quantity. The exponents $ x_{1},x_{2},x_{3}$ and $x_{4}$ are unknown, and our main objective is to determine their numerical values. Starting from the rules of dimensional algebra, and remembering that the notation $[x]$ means “the dimension of $x$” [11], we can rewrite Eq. (4) as:

\begin{equation} 
\left[ T_{H} \right] = \left[ \Pi \right] \left[ c \right]^{x_{1}} \left[ \hbar \right]^{x_{2}} \left[ k \right]^{x_{3}} \left[ \mu \right]^{x_{4}}. 
\end{equation}

Since $\Pi$ is dimensionless, it must be true that $[\Pi]=1$. By introducing the dimensions of the five parameters listed in Table 3, we obtain:

\begin{equation} 
M^{0}L^{0}T^{0}\Theta^{1} = \left( M^{0}L^{1}T^{-1}\Theta^{0}\right)^{x_{1}}   \left(  M^{1}L^{2}T^{-1}\Theta^{0}\right)^{x_{2}}  \left(  M^{1}L^{2}T^{-2}\Theta^{-1}\right)^{x_{3}}  \left(  M^{0}L^{3}T^{-2}\Theta^{0} \right)^{x_{4}}. 
\end{equation}

After sorting and grouping terms, we have:

\begin{equation} 
M^{0}L^{0}T^{0}\Theta^{1} = M^{0x_{1} + x_{2} + x_{3} + 0x_{4}} L^{x_{1} + 2x_{2} + 2x_{3} + 3x_{4}} T^{-x_{1} -x_{2} - 2x_{3} -2x_{4}} \Theta^{0x_{1} + 0x_{2} - x_{3} + 0x_{4}}. 
\end{equation}

To satisfy this equality, each exponent on the right side must be equal to the corresponding exponent on the left side, producing a linear system of four equations with four unknowns:

\begin{equation} 
  \systeme{
  0x_{1}+x_{2}+x_{3}+0x_{4} = 0, 
  x_{1}+2x_{2}+2x_{3}+3x_{4} = 0, 
  -x_{1}-x_{2}- 2x_{3}-2x_{4} = 0, 
  0x_{1}+ 0x_{2}-x_{3}+ 0x_{4} = 1
  }.
\end{equation}

By substitution, the reader can verify that the solution of this system is $x_{1}=3, x_{2}=1, x_{3}=-1, x_{4}=-1$. Substituting these values into Eq. (4) gives

\begin{shaded}

\begin{equation} 
T_{H} =\Pi \frac{\hbar c^{3}}{k \mu} = \Pi \frac{\hbar c^{3}}{kGM_{BH}}.
\end{equation}

\end{shaded}

Except for the dimensionless parameter $\Pi$, this equality is identical to Eq. (1). Comparing Eqs. (1) and (9) we see that $\Pi = 1/8\pi$. The interested reader can find other simple derivations of the Hawking temperature in [2,12].

\section{The physical meaning of Hawking's discovery: The Hawking effect}

Hawking temperature is not an isolated result, as it has two physical implications that continue to raise questions: the first is that black holes evaporate, and the second is that they have entropy. To understand these two important results in general terms, we must begin by exploring the physical meaning of the Hawking temperature. 

\begin{table}[htbp] 
\begin{center}
\caption{The Hawking effect}
\resizebox{1\textwidth}{!} {
\begin{tabular}{ m{4cm}  m{6cm}  m{10cm} }
\toprule
\textbf{Discovery } & \textbf{Mathematical formulation} & \textbf{Description} \\
\midrule
Hawking temperature & $T_{H} = \frac{\hbar c^{3}}{8\pi kG M_{BH}}$ & Black holes have an absolute temperature $T_{H}$ that depends inversely on their mass $M_{BH}$. The greater the value of $M_{BH}$, the lower the value of$T_{H}$, and vice versa. For known black holes, $T_{H}$ is undetectable. \\ 

Hawking radiation and evaporation & $t_{ev} \cong 10^{3} \frac{G^{2}M_{BH}^{3}}{c^{4}\hbar}$ & Black holes evaporate in a characteristic time $t_{ev}$ that depends on the cube of their mass. For known black holes, $t_{ev}$ is much older than the age of the universe, and evaporation is undetectable. \\ 

Bekenstein-Hawking entropy  & $S_{BH} = \frac{kc^{3}A_{BH}}{4G\hbar}=\frac{4\pi kGM_{BH}^{2}}{\hbar c}$ & Black holes have an entropy $S_{BH}$  that depends on the horizon area $A_{BH}$, which is proportional to the square of the mass. The maximum entropy that can accumulate in a region of fixed size is that of the largest black hole confined in said region.  \\ 
\midrule
\end{tabular}
}
\label{The Hawking effect}
\end{center}
\end{table}

Table 4 summarises the equations that describe the \textit{Hawking effect}, which is the name given to the set of phenomena associated with Hawking's discovery. The interested reader will find very simple derivations of these equations in [2]. In the following sections, we will delve into the Hawking effect. Due to space limitations, we will not analyse the physical mechanism responsible for the Hawking temperature. A reader interested in this topic can refer to Hawking's texts [7,8,16], or other works published in this journal, where this topic is addressed in an accessible way [2,12].

\subsection{Hawking temperature}

The first aspect that we must keep in mind in relation to the Hawking temperature is that it does not depend on the interaction of the black hole with the matter in its environment. A black hole could be completely isolated and still have a temperature given by Eq. (1).

We will start by rewriting this equation in a form that allows for quick calculations, using the values of $G,c,\hbar,k$ from Table 1. If we enter the solar mass $M_{\odot}=1.99 \times 10^{30} kg$, and multiply the right side of Eq. (1) by the dimensionless quotient $M_{\odot}/1.99 \times 10^{30} kg=1$, we obtain the Hawking temperature expressed in kelvin ($K$) as:

\begin{equation} 
T_{H} = 6.17 \times 10^{-8}K\left( \frac{M_{\odot}}{M_{BH}} \right), 
\end{equation}

where $T_{H}$ is the temperature of an isolated black hole, that is, a black hole where the outer region of its horizon contains no matter or radiation. This equation tells us that “black holes ain’t so black”, and that they can be very bright and hot if $M_{BH}$ is small enough. To provide empirical corroboration of Hawking's discovery, we need to find light black holes. The lightest ones that have been observed so far have masses of the order of $\sim M_{\odot}$, while the heaviest ones can have masses of up to $\sim 10^{10}M_{\odot}$ [13]. To obtain an upper bound for $T_{H}$, we take $M_{BH} = M_{\odot}$ in Eq. (10), obtaining $T_{H} = 6.17 \times 10^{-8} K$. This tiny temperature is undetectable through astronomical observations.

\begin{figure}[h]
  \centering
    \includegraphics[width=0.55\textwidth]{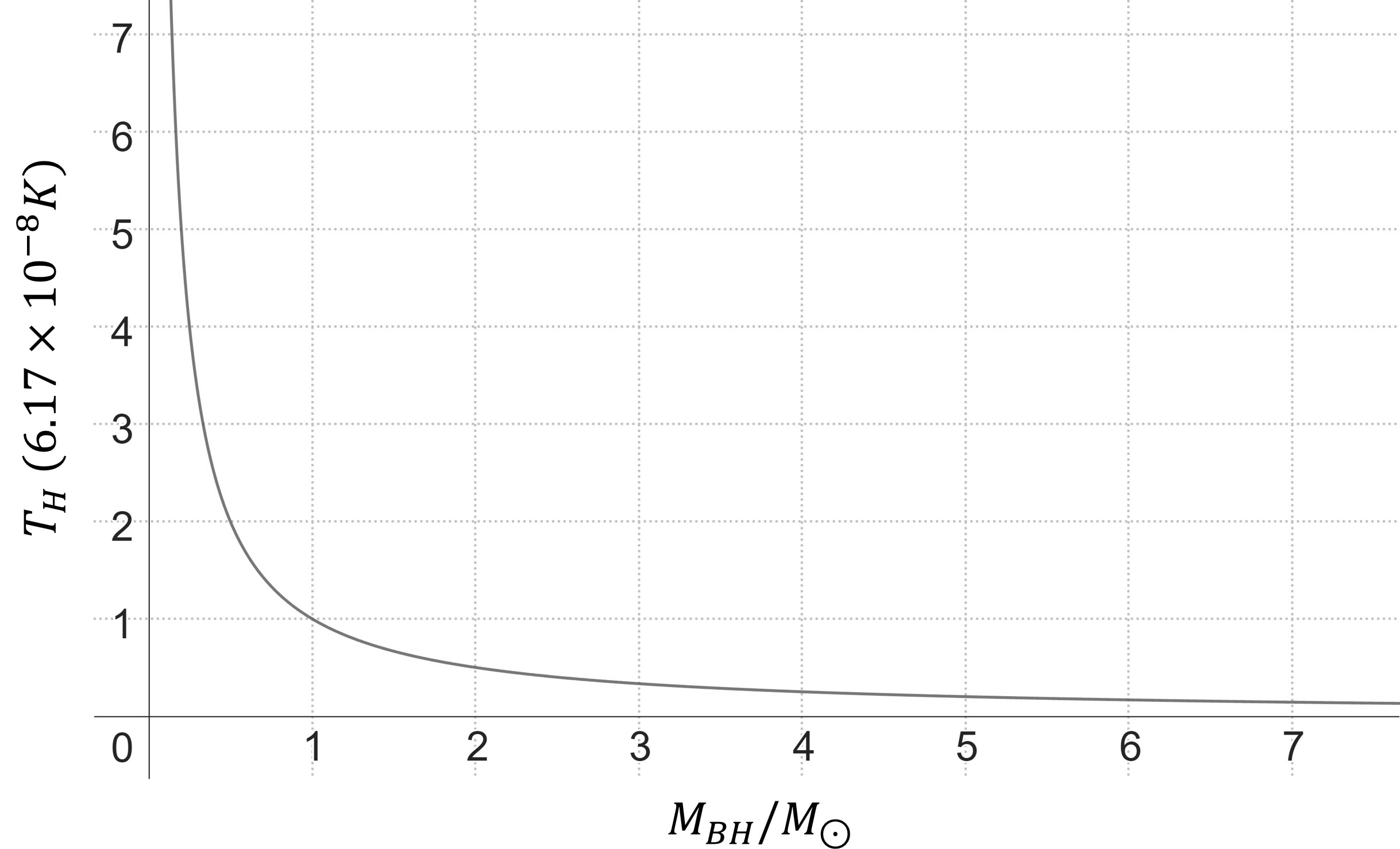}
  \caption{Graph of Hawking temperature, $T_{H}$, versus $M_{BH}/M_{\odot}$.}
\end{figure}

To illustrate the relationship between $T_{H}$ and $M_{BH}$ more clearly, a graph of Eq. (10) is shown in Fig. 4. The horizontal axis has dimensionless units of $M_{BH}/M_{\odot}$ and the vertical axis has units of $6.17\times 10^{-8} K$. It can be observed that $T_{H}$ approaches zero asymptotically as $M_{BH}$ increases.

\subsection{Hawking radiation and evaporation}

As with any hot body, a black hole emits mainly photons; that is, Hawking radiation is basically light. However, if $T_{H}$ is high enough, it can also emit neutrinos, electrons, positrons, etc. This emission process is known as evaporation [14]. Just as the evaporation of a puddle of water on a summer day gradually reduces the mass of the puddle, Hawking radiation gradually reduces the mass of the black hole, as the photons, electrons, and other particles cross the horizon to the outside.\\

Eq. (10) reveals that the smaller the value of $M_{BH}$, the more Hawking radiation is emitted from the horizon, and the more radiation is emitted, the more rapidly $M_{BH}$ is reduced. This means that evaporation does not occur at a constant rate, but instead increases over time [7,15,16]. However, since black holes are not isolated, under realistic conditions it is only when they emit more material than they absorb that there will there be a net outward flow of radiation that will reduce $M_{BH}$. The time taken for the mass of an isolated black hole of initial mass $M_{BH}$ to be reduced to zero is called the evaporation time [17]:

\begin{equation}
t_{ev} \cong 10^{3} \frac{G^{2}M_{BH}^{3}}{c^{4}\hbar}.
\end{equation}

If we introduce the values of $G,c,\hbar$ and multiply the right side of this equality by the dimensionless quotient $(1.99\times 10^{30} kg /M_{\odot} )^{3}=1$, the evaporation time is expressed in seconds ($s$) as:

\begin{equation}
t_{ev} = 10^{73}s \left( \frac{M_{BH}}{M_{\odot}} \right)^{3}. 
\end{equation}

Fig. 5 shows a graph of this equation, where the horizontal axis has units of $M_{BH}/M_{\odot}$ and the vertical axis has units of $10^{73} s$. It can be observed that $t_{ev}$ increases rapidly with an increase in $M_{BH}$.\\

Eq. (12) reveals that, in general, $t_{ev}$ is extraordinarily large. To obtain a lower bound for $t_{ev}$, we take $M_{BH}=M_{\odot}$ in Eq. (12), obtaining $t_{ev} \cong 10^{73} s$. This figure is 56 orders of magnitude greater than the age of the universe, which is estimated at $\sim 10^{17}s$. Hence, evaporation is unobservable, which is consistent with the fact that $T_{H}$ is undetectable.

\begin{figure}[h]
  \centering
    \includegraphics[width=0.55\textwidth]{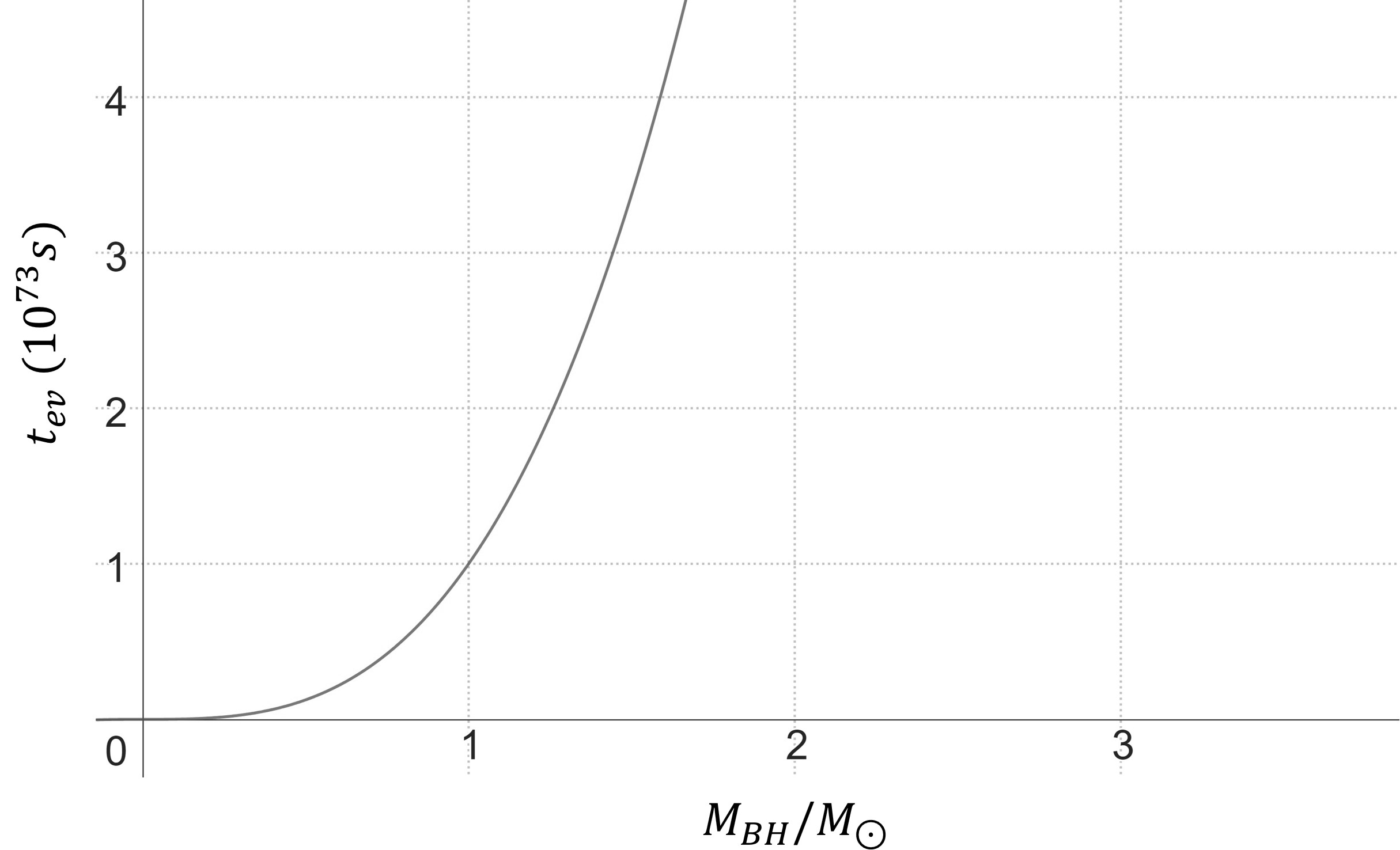}
  \caption{Graph of evaporation time, $t_{ev}$, versus  $M_{BH}/M_{\odot}$.}
\end{figure}

\subsection{Bekenstein-Hawking entropy}

The laws of thermodynamics establish that every body with a non-zero temperature has entropy, and black holes are no exception. According to calculations originally carried out by the physicist Jacob Bekenstein and perfected by Hawking [1,18], the \textit{Bekenstein-Hawking entropy} $S_{BH}$ of an isolated black hole is proportional to the area of its event horizon:

\begin{equation}
S_{BH} = \frac{kc^{3}A_{BH}}{4G\hbar}.
\end{equation}

Since $A_{BH}$ is given by Eq. (3), the entropy can also be written as a function of the mass of the black hole and the solar mass:

\begin{equation}
S_{BH} = \frac{4\pi kGM_{BH}^{2}}{\hbar c} \cong 1.5 \times 10^{54} J\cdot K^{-1} \left( \frac{M_{BH}}{M_{\odot}} \right)^{2}, 
\end{equation}

where we have introduced the values of the constants and multiplied the right side of the equality by the dimensionless quotient $(1.99\times 10^{30} kg /M_{\odot} )^{2}=1$, allowing $S_{BH}$ to be expressed in joules/kelvin ($J\cdot K^{-1}$), which are the SI units of entropy. Fig. 6 shows a graph of this equation, where the horizontal axis has units of $M_{BH}/M_{\odot}$ and the vertical axis units of $1.5\times 10^{54} J \cdot K^{-1}$. To obtain a lower bound on $S_{BH}$, we take $M_{BH}=M_{\odot}$ in Eq. (14), which gives $S_{BH}=1.5\times 10^{54} J \cdot K^{-1}$. This is a colossal number, and reveals that black holes are among the most entropic objects in the universe.

\begin{figure}[h]
  \centering
    \includegraphics[width=0.55\textwidth]{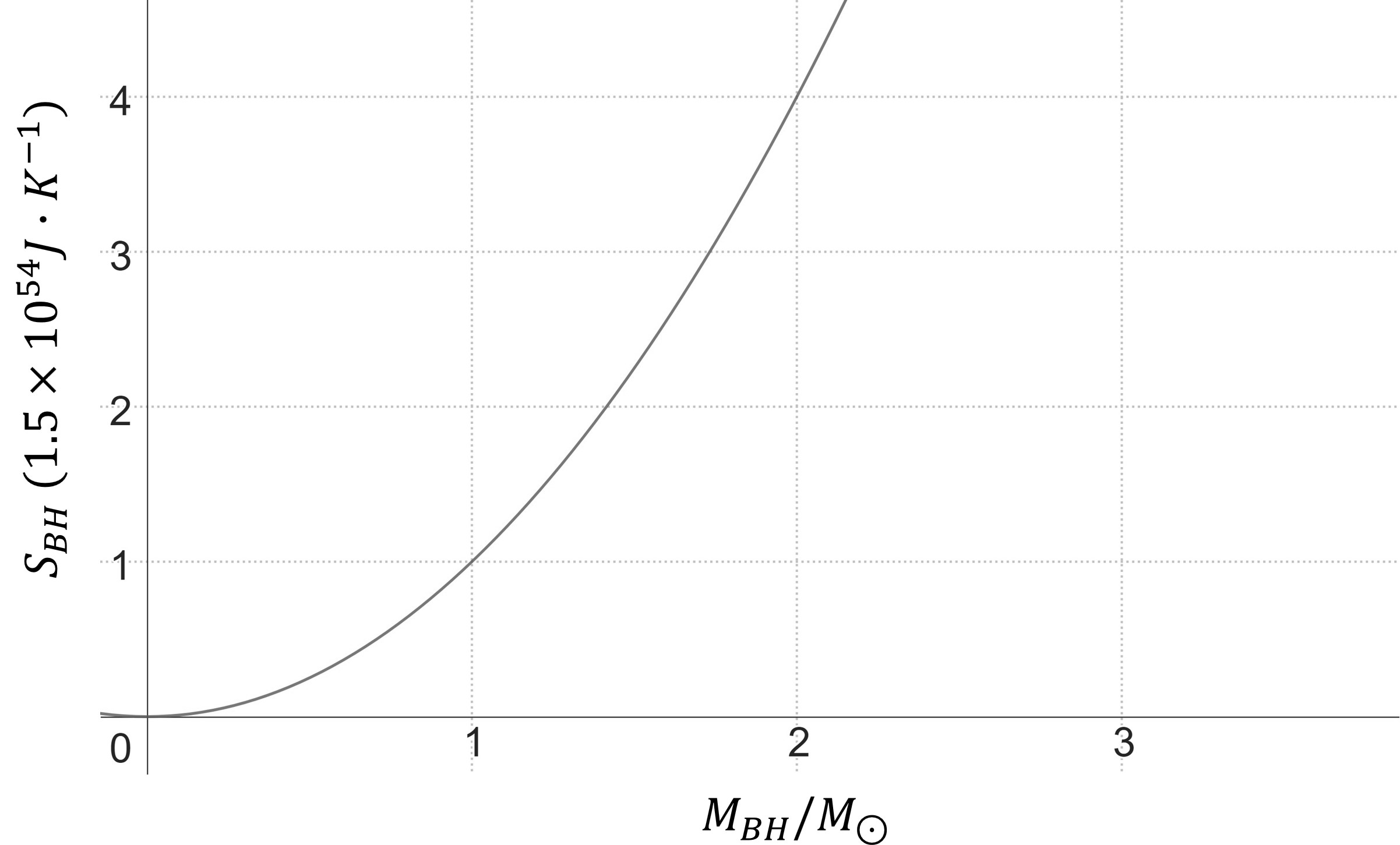}
  \caption{Graph of entropy, $S_{BH}$, versus  $M_{BH}/M_{\odot}$.}
\end{figure}

\section{Analogue gravity}

If, as the evidence suggests, the smallest black holes have masses on the order of $\sim M_{\odot}$, with Hawking temperatures on the order of a millionth of a kelvin, attempts to detect Hawking radiation may be unsuccessful. To get around this difficulty, the ideal would be to create black holes in our laboratories, which would allow us to study them in detail, but this is obviously impossible with current technology [19], and it may never be possible to develop such technology.\\ 

Due to these difficulties, some specialists have turned their attention towards a field of experimental physics called \textit{analogue gravity}, whose objective is implementing physical systems that imitate the behaviour of gravitational phenomena that are difficult to observe under real conditions [20], such as the Hawking effect.\\ 

The first person to propose the use of analogue models to study black holes was the Canadian theoretical physicist William Unruh, who, in 1972, introduced an inventive analogy: he imagined a fish screaming while falling over a waterfall. As the fish falls, it heads toward a point in the waterfall where the water travels downward faster than sound waves can travel upward. When the fish crosses this point, its friends in the river above can no longer hear it, since to enable this, the speed of the sound waves would have to exceed the speed of sound in the water (Fig. 7). According to Unruh, this is analogous to what would happen to an astronaut who fell into the jaws of a black hole: the astronaut would not be able to communicate with the base station on Earth, because the signals sent towards the outside would need to move faster than light, which is prohibited by physical laws. \\

\begin{figure}[h]
  \centering
    \includegraphics[width=0.35\textwidth]{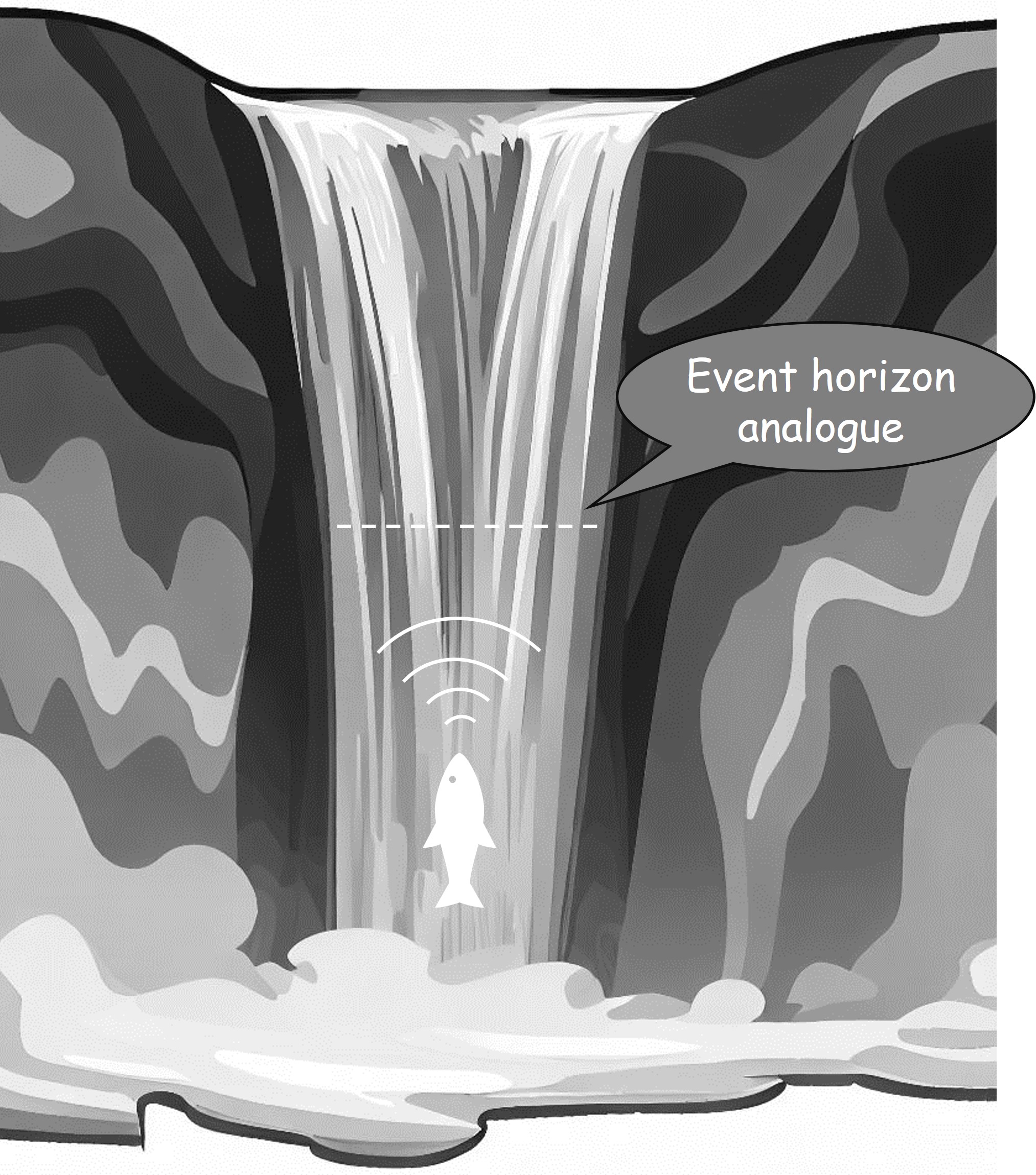}
  \caption{A fish sends sound waves up a waterfall. The dashed line is an analogue of the horizon, where the speed with which the waterfall falls is equal to the speed of sound in the water, meaning that the waves cannot cross the dashed line.}
\end{figure}

Almost a decade later, Unruh showed that the event horizons of black holes and the sonic horizons in systems such as his waterfall, which are now known as \textit{sonic black holes}, can be described by identical equations [21]. In particular, Unruh showed that sonic black holes can mimic the Hawking temperature and Hawking radiation. The results of the first experiment of this type were published in 2014 [20], so this is a new and developing field of research. The results obtained so far are encouraging, although not conclusive, and the evidence is consistent with the discoveries made by Hawking half a century ago.

\section{Final comments: Epitaph for a genius}

The great discoveries of physics are expressed through equations that accurately describe a wide range of phenomena. Some equations stand out for their simplicity, others for their elegance, and others for their ability to build bridges between apparently dissimilar domains of reality. However, there is a select group of equations that stand out for combining all these virtues.\\

For example, Newton's law of gravitation falls into this category, as it reconciles celestial physics with terrestrial physics, as does Einstein's mass-energy equivalence, which unifies matter and energy. The Hawking equation also falls into this category, as it bridges general relativity, quantum mechanics and thermodynamics. Hawking's tomb in Westminster Abbey has this equation engraved on it, alongside the inscription: “Here lies what was mortal of Stephen Hawking”. It is hoped that this work will contribute to a better understanding of Hawking's work and his immortal equation, a symbol of a scientific revolution that is far from over.

\section*{Appendix: Dimensional analysis and dimensional algebra}

A \textit{physical dimension}, briefly called a \textit{dimension}, is a property of an object or phenomenon that can be measured and expressed in terms of standard fundamental units. Dimensional analysis is a tool used in science and engineering to study and relate different physical quantities through their dimensions. For example, length, mass, and time are dimensions that have associated SI units of the metre, kilogram, and second, respectively. There are seven dimensions, which correspond to the seven fundamental physical magnitudes. These dimensions, their symbols, and the associated SI units appear in Table A1 below.\\

In the various applications of dimensional analysis, the notation $[a]$ is commonly used, which is read as “the dimensions of $a$”. For example, if $m$ is the mass of any object, $\rho$ is its density and $v$ is its speed, we have $[m]=M$, $[\rho]=ML^{-3}$ and $[v]=LT^{-1}$.

\begin{table}[htbp] 
\begin{center}
\caption*{\textbf{Table A1}: SI units and their corresponding dimensions}
\resizebox{0.6\textwidth}{!} {
\begin{tabular}{l l l l} 
\toprule
\textbf{Dimension } & \textbf{Symbol} & \textbf{SI unit} & \textbf{SI symbol} \\
\midrule
Mass & $M$ & kilogram & $kg$ \\ 

Lenght & $L$ & metre & $m$ \\ 

Time & $T$ & second & $s$  \\ 

Temperature & $\Theta$ & kelvin  & $K$  \\

Electric current & $I$ & ampere  & $A$  \\

Luminous intensity & $J$ & candle  & $Cd$  \\

Amount of substance & $N$ & mole  & $mol$  \\

\midrule
\end{tabular}
}
\label{SI units and their corresponding dimensions}
\end{center}
\end{table}

To apply dimensional analysis, it is useful to systematise seven principles that, together, we can call the \textit{rules of dimensional algebra}. We can consider a set of quantities $a,b,c...$ that may be pure numbers, variables, physical constants, etc. The rules of dimensional algebra that can be applied to these quantities are as follows:

\begin{enumerate}[label= \roman*]

\item \textit{Product rule}: $[ab]=[a][b]$.
\item \textit{Quotient rule}: $[a/b]=[a]/[b]$.
\item \textit{Commutativity rule}: $[a][b]=[b][a]$.
\item \textit{Associativity rule}: $[a]([b][c])=([a][b])[c]$.
\item \textit{Powers rule}: $[a^{n}]=[a]^{n}$.
\item \textit{Homogeneity rule}: If $a\pm b=c$ then $[a\pm b]=[a]=[b]=[c]$.
\item \textit{Dimensionless quantities rule}: If $a$ is dimensionless, then $[a]=1$.

\end{enumerate}

\section*{Acknowledgments}
I would like to thank to Daniela Balieiro for their valuable comments in the writing of this paper. 

\section*{References}

[1] S.W. Hawking, Particle creation by black holes, Communications in Mathematical Physics. 43 (1975) 199–220.

\vspace{2mm}

[2] J. Pinochet, “Black holes ain’t so black”: An introduction to the great discoveries of Stephen Hawking, Phys. Educ. 54 (2019) 035014.

\vspace{2mm}

[3] C.F. Bohren, Dimensional analysis, falling bodies, and the fine art of not solving differential equations, American Journal of Physics. 72 (2004) 534–537. https://doi.org/10.1119/1.1574042.

\vspace{2mm}

[4] A. Einstein, Die Grundlage der allgemeinen Relativitätstheorie, Annalen Der Physik. 354 (1916) 769–822.

\vspace{2mm}

[5] J. Pinochet, Five misconceptions about black holes, Phys. Educ. 54 (2019) 55003.

\vspace{2mm}

[6] B. Schutz, Gravity from the Ground Up, Cambridge University Press, Cambridge, 2003.

\vspace{2mm}

[7] S.W. Hawking, A brief history of time, Bantam Books, New York, 1998.

\vspace{2mm}

[8] S.W. Hawking, The quantum mechanics of black holes, Scientific American. 236 (1976) 34–40.

\vspace{2mm}

[9] S.W. Hawking, Black Hole explosions?, Nature. 248 (1974) 30–31.

\vspace{2mm}

[10] T. Szirtes, Applied Dimensional Analysis and Modeling, 2nd ed., Elsevier, Oxford, 2007.

\vspace{2mm}

[11] D.S. Lemons, A Student’s Guide to Dimensional Analysis, Cambridge University Press, Cambridge, 2017.

\vspace{2mm}

[12] J. Pinochet, Three easy ways to the Hawking temperature, Physics Education. 56 (2021) 053001. https://doi.org/10.1088/1361-6552/AC03FC.

\vspace{2mm}

[13] K.L. Lang, Essential Astrophysics, Springer, Berlin, 2013.

\vspace{2mm}

[14] R.J. Adler, P. Chen, D.I. Santiago, The Generalized Uncertainty Principle and Black Hole Remnants, General Relativity and Gravitation. 33 (2001) 2101–2108.

\vspace{2mm}

[15] S.W. Hawking, Black Holes and Thermodynamics, Physical Review D. 13 (1976) 191–197.

\vspace{2mm}

[16] S.W. Hawking, The Universe in a Nutshell, Bantam Books, New York, 2001.

\vspace{2mm}

[17] D.N. Page, Particle emission rates from a black hole: Massless particles from an uncharged, nonrotating hole, Physical Review D. 13 (1976) 198–206.

\vspace{2mm}

[18] J.D. Bekenstein, Black Holes and Entropy, Physical Review D. 7 (1973) 2333–2346.

\vspace{2mm}

[19] https://www.scientificamerican.com/article/has-anyone-created-a-black-hole-on-earth/

\vspace{2mm}

[20] J. Steinhauer, Observation of self-amplifying Hawking radiation in an analogue black-hole laser, Nature Physics. 10 (2014) 864–869.

\vspace{2mm}

[21] W.G. Unruh, Experimental Black-Hole Evaporation?, Physical Review Letters. 46 (1981) 1351–1353.

\end{document}